\documentclass[traditabstract]{aa} 
\usepackage{graphicx}
\usepackage{txfonts}
\usepackage{wasysym}
\usepackage{natbib}
\bibpunct{(}{)}{;}{a}{}{,} 
\usepackage{ulem}
\begin{document}
   \title{
TASTE. II. A new observational study \\ of transit time variations in 
HAT-P-13b\thanks{Based on observations collected at Asiago observatory.}}
   \author{V. Nascimbeni\inst{1,2,4}\thanks{Visiting PhD Student at STScI (DDRF D0001.82432
           program).}
          \and G. Piotto\inst{1,4}
          \and L.\ R. Bedin\inst{2} \and M. Damasso\inst{1,3} \and L. Malavolta\inst{1}
          \and L. Borsato\inst{1,4}
          }

   \institute{Dipartimento di Astronomia, Universit\`a degli Studi di Padova,
              Vicolo dell'Osservatorio 3, 35122 Padova, Italy\\
              \email{valerio.nascimbeni@unipd.it, giampaolo.piotto@unipd.it}
         \and
             Space Telescope Science Institute,
             3700 San Martin Drive, Baltimore, MD 21218\\
             \email{bedin@stsci.edu}
         \and
             Astronomical Observatory of the Autonomous Region of the Aosta Valley,
             Loc.\ Lignan 39, 11020 Nus (AO), Italy
         \and
             INAF -- Osservatorio Astronomico di Padova, vicolo dell'Osservatorio 5, 35122 Padova, Italy
             }

   \date{Submitted March 3, 2011; Accepted May 22, 2011}

\abstract{ TASTE (The Asiago Search for Transit timing variations of
  Exoplanets) project is collecting high-precision, short-cadence
  light curves for a selected sample of transiting exoplanets.  
  It has been claimed that the
  hot jupiter HAT-P-13b suddenly deviated
  from a linear ephemeris by $\sim 20$ min, 
  implying that there is
  a perturber in the system.  Using five new transits, we discuss the
  plausibility of this transit time variation (TTV), and show that a
  periodic signal should not be excluded. More follow-up observations
  are required to constrain the mass and the orbit of the hypothetical
  perturber.  }

   \authorrunning{Nascimbeni et al.}
   \titlerunning{A new observational study of transit time variations in HAT-P-13b}
   \maketitle

\section{Introduction}

Photometric transits represent a great opportunity to discover and
characterize extrasolar planets. They are, for instance, the only
direct method to estimate the planetary radius and to constrain other
important physical and orbital parameters \citep{winn2010a}.  In
principle, a single planet orbiting the host star in a Keplerian orbit
is expected to transit at strictly periodic time intervals, unless it
is perturbed by a third body \citep{holman2005}.  By performing
accurate measurements of the central instant time of a known
transiting planet, it would be possible to detect deviations from a
linear ephemeris, and to infer the parameters of the perturber
\citep{agol2005}.  Such a search for other bodies via transit time
variations (TTV) is very sensitive to low-mass planets when they are
locked in low-order orbital resonances. In these orbits, even
earth-mass perturbers would cause TTVs of the order of a few minutes,
i.e.  easily detectable with ground-based techniques.

In the past few years, some authors have claimed to have detected TTVs
using ground-based facilities, for instance from WASP-3b \citep{mac2010},
WASP-10b \citep{mac2010b}, and WASP-5b \citep{fukui2010}, though none
have been confirmed so far. In contrast, the Kepler mission found
undisputable mutual TTVs for the double transiting system Kepler-9b,c
\citep{holman2010} and for five among six planets transiting on
Kepler-11 \citep{lissauer2011}, which has lead to the validation of those
planets, as well as a deep characterization of their planetary
systems.

\citet{pal2011} claimed to have detected an unusually
large TTV in HAT-P-13b.  The G4V star HAT-P-13 hosts a multiple
planetary system, and was the first multiple system discovered with a
transiting planet. HAT-P-13b is a classical ``hot jupiter'' ($M=0.85
M_j$, $R=1.28 R_j$) transiting every $\sim 2.91$ days, while HAT-P-13c
is an outer, massive companion ($M\sin i\sim15 M_j$, $P=428.5$ days)
detected only with radial velocity (RV) measurements
\citep{bakos2009}. A 2010 multi-site campaign designed to detect the
transit of HAT-P-13c yielded a null result with a 65-72\% significance
level \citep{szabo2010}.  A long-term RV trend of HAT-P-13 was
observed by \citet{winn2010b} and interpreted as evidence of a third
companion with an even longer orbital period, still to be constrained.

\begin{table*}
\caption{Summary of the observed transits of HAT-P-13b at the Asiago 1.82m telescope.}
\label{observ}
\centering
\begin{tabular}{llllllll}
\hline\hline
eve. date   & UT obs. time  & airmass            & exptime (s)  & cadence (s)  & duty-cycle & frames & notes      \\ \hline
2011 Jan 2  & 19:46$-$23:40 & 1.63$-$1.03        & 8            & 9.7          & 83\%       & 1451   & clouds     \\
2011 Jan 31 & 23:08$-$04.18 & 1.00$-$1.58        & 3, 5         & 6.6          & 75\%       & 2810   & clear \\
2011 Feb 3  & 21:07$-$03:30 & 1.05$-$1.00$-$1.23 & 5            & 6.7          & 75\%       & 2902   & some veils \\
2011 Feb 6  & 18:53$-$23:40 & 1.29$-$1.00$-$1.01 & 4, 5         & 5.8          & 71\%       & 2940   & clear \\
2011 Feb 9  & 17:29$-$22:31 & 1.57$-$1.00        & 4, 5         & 6.0          & 72\%       & 3011    & clear, twilight \\ \hline
\end{tabular}
\tablefoot{The columns give: the ``evening date'' of the observation, the UT time span of the photometric series,
the airmass evolution, the exposure time and the average net cadence in s, the overall duty-cycle, the
number of frames gathered, and the sky conditions at the time.}
\end{table*}

The TTV claimed by \citet{pal2011} appears to be a sudden deviation of
the timings of three transits (by 3.3, 5.5, 8.4 $\sigma$) from the
linear ephemeris evaluated using the previous data.  All the three
newly added points are consistent with each other.
The ``switch'' has an amplitude of the
order of $\sim 0.015$ days (Fig. \ref{ttvsinus}, top left panel).  
This would make it the largest TTV claimed from the
ground. The presence of the outer companion HAT-P-3c does not explain
such a perturbation, as its expected TTV would have an amplitude of a
few seconds and a $\sim 430$ d period, while the measurements before
2011 are in agreement with a constant ephemeris. An intriguing
possibility is that this behaviour is induced by a long-period,
massive companion on a very eccentric orbit. Eccentric perturbers are
known to cause sudden ``spikes'' in an otherwise constant $O-C$
diagram \citep{holman2005}.

Unfortunately, the available data allow us to constrain the
orbital parameters of neither the hypotetical perturber, nor its mass.  The
transit of HAT-P-13b is shallow ($\Delta m\sim0.008$ mag) and long
($d\sim194$ min), i.e. very difficult to monitor. Few measurements 
were made before 2011, and most with an estimated timing
accuracy $\gg1$ min.  Two out of three transits from \citet{pal2011}
show a considerable amount of systematic errors, and one is partial,
lacking the egress. Their detection urgently needs a confirmation: if
confirmed, efforts should be made to monitor other transits of
HAT-P-13b in the near future, to assess the parameters and
the nature (substellar or planetary) of the perturber.

In this Paper, we report five new high-precision light curves of
HAT-P-13b, observed in January and February 2011 with the Asiago 1.82m
telescope. Four of those transits are consecutive, and their estimated
timing accuracy is the highest ever achieved for this target. We confirm
the deviation with respect to the ephemeris from \citet{szabo2010} reported by
\citet{pal2011}. No ephemeris with a constant period can be fitted to
the data with an acceptable $\chi^2$.  The observed deviation 
is still highly unconstrained.  We note that a
long-period, sinusoidal TTV can be fitted to the $O-C$ points,
with only one significant outlier.

\section{Observations}

All the observations reported here were made as part
of the TASTE (The Asiago Search for Transit timing variations of
Exoplanets) project \citep{nascimbeni2010}. TASTE is collecting
high-precision, short-cadence light curves for a selected sample of
transiting exoplanets, to discover low-mass planetary
companions or exomoons with the TTV/TDV method (transit time/duration
variation). We refer to that paper for a detailed description of our
instrumental setup, observing strategy, and data
reduction/analysis. HAT-P-13b is among the sample we are following.

We collected five transit light curves of HAT-P-13b using the AFOSC
imager with its new E2V 42-20 CCD detector mounted at the Asiago 1.82m
telescope\footnote{\sf http://www.pd.astro.it/asiago/}.  An
observation log is shown in Table \ref{observ}. All the observations
were made using a standard Cousins $R$ filter and $4\times4$
binning. We employed binning and windowing to speed up the readout and
decrease as much as possible the technical ``dead'' times between the
exposures. We achieved an average $>70\%$ duty-cycle and a $<10$ s net
cadence for all our photometric series.  We acquired both sky- and dome
flat-field frames during each night; bias and dark frames were taken at
both the beginning and the end of a light curve to constrain possible
instrumental drifts.

Stellar profiles were defocused to $\sim4-6''$ FWHM (that is, over
$\sim1300$ physical pixels) in order to minimize systematic errors
arising from imperfect flat-field correction, guiding drifts, and
pixel-to-pixel inhomogeneity. The $9'\times 2'.6$ CCD window that we read
included HAT-P-13 as well as the main reference star TYC 3416-1608-1,
a star with a magnitude and colour similar to HAT-P-13 ($V_T=10.80$
versus (vs.) 10.50 and $B_T-V_T=0.81$ vs. 0.52).

\section{Data reduction and analysis}

\begin{figure*}
\centering
\includegraphics[height=17cm,trim=30 30 0 00]{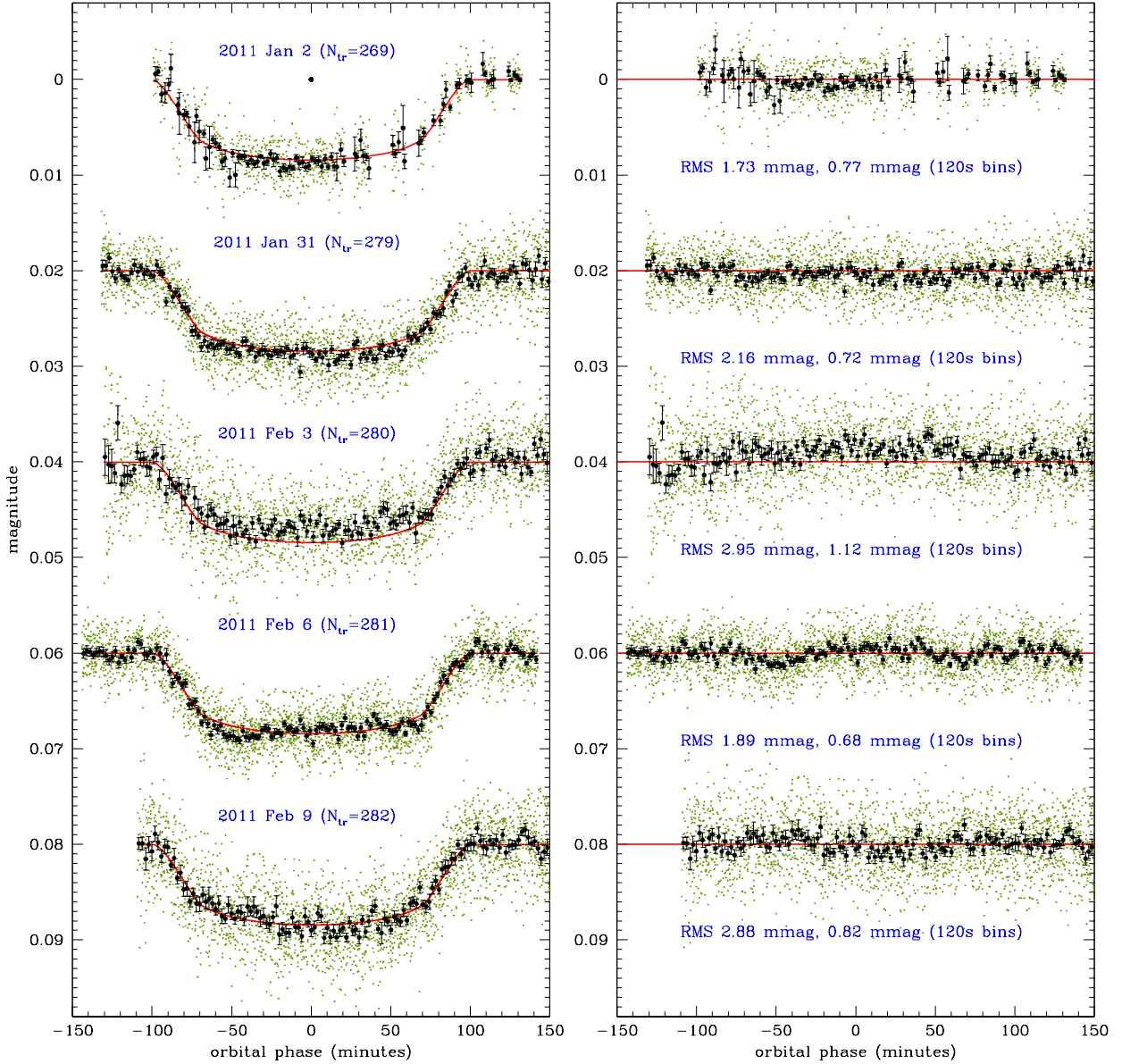}
\caption{
(\emph{Left}):
Light curves of HAT-P-13b taken at the Asiago 1.82m telescope, for the five
transit summarized in Table \ref{timings}. The unbinned photometric points
are plotted in green, the 120-s binned points are plotted in black. The red line
is the best-fit model fitted by JKTEBOP. Transits have been offset by
intervals of 0.02 mag for clarity.
(\emph{Right}): photometric residuals around the best-fit model.}
\label{lcurves}
\end{figure*}

We performed differential aperture photometry on HAT-P-13 using
STARSKY \citep{nascimbeni2010}, an independent pipeline that we
specifically developed for the TASTE project. This code is designed to
keep under control any possible source of systematic errors, and
implements a fully empirical, iterative approach to identify and
correct them. The output light curve is the one with the smallest
effective RMS. Specific diagnostics are evaluated at each iteration to
constrain the amount of correlated noise. The final, detrended light
curves are shown in Fig. \ref{lcurves}, both unbinned and binned over
120 s intervals.  The photometric RMS scatter is in the range $\sigma_u
= 1.7-2.9$ mmag for the unbinned points and the range $\sigma_{120} =
0.6-1.1$ mmag for the 120 s bins.  Three of the light curves in Fig.~1
represent the most accurate light curves of HAT-P-13b published so
far.

We ran the JKTEBOP code version 25 \citep{southworth2004} to fit a
transit model over our light curves. We used a quadratic law for limb
darkening, fixing both the linear and the quadratic term $u_1$, $u_2$
to the theoretical values interpolated from the \citet{claret2000}
tables, for the stellar parameters of HAT-P-13 derived by
\citet{bakos2009}. Three of the remaining parameters of the transit
(inclination $i$, ratio, and sum of the fractional radii $R_a/R_b$,
$R_a+R_b$)  were estimated by fitting  the two highest quality light curves (2011
Jan 3 and Feb 6). We then fixed $i$, $R_a/R_b$, $R_a+R_b$ to these respective
values, and fitted each individual transit only for the central instant
$T_0$. Since the formal errors derived by the least squares routine are
known to be far too optimistic, we took advantage of two techniques
implemented in JKTEBOP to estimate realistic errors: a Monte Carlo
test (MC) and a bootstrapping method based on the cyclic permutations
of the residuals (RP or ``prayer bead'' algorithm,
\citealt{southworth2008}). The errors from the RP algorithm are
significantly larger, suggesting a non-negligible amount of red noise
in our light curves.  We therefore adopted conservatively the RP
1-$\sigma$ errors in our analysis. The best-fit $T_0$ for each
transit, converted from UT to barycentric Julian date (BJD), are shown
in Table \ref{timings} along with their estimated uncertainties.

\begin{table}
\caption{Best-fit values of the central instant $T_0$ for the five reported new transits of HAT-P-13b.}
\label{timings}
\centering
\begin{tabular}{lllll}
\hline\hline
$N_\textrm{tr}$ & BJD $T_0$ (LS) & BJD $T_0$ (RP) & $\Delta T_0+$ & $\Delta T_0-$ \\ \hline
269 & 2455564.39839  & 2455564.39892 & 0.00089 & 0.00271 \\
279 & 2455593.56110  & 2455593.56085 & 0.00114 & 0.00115 \\
280 & 2455596.47625  & 2455596.47610 & 0.00299 & 0.00311 \\
281 & 2455599.39230  & 2455599.39252 & 0.00046 & 0.00105 \\
282 & 2455602.31031  & 2455602.31038 & 0.00167 & 0.00166 \\ \hline

\end{tabular}
\tablefoot{The columns give: the ``event number'' $N_\textrm{tr}$ for the transit following the
ephemeris by \citet{bakos2009}, the central instant of the transit $T_0$
as estimated by a simple least squares fit (LS) and
by the residual-permutation technique (RP), and the associated 1-$\sigma$
uncertainties $\Delta T_0$ (in days) as given by the RP distribution. BJD times are 
calculated from UTC.}
\end{table}

\section{Discussion}

Our five timings points are very close in time to each other,
four of them being consecutive transits
($N_{\textrm{tr}}=258,259,260,261$).  They are consistent within the
RP errors with a constant ephemeris ($O-C$ = $+88$, $+0$, $-89$,
$-79$, $+55$ s) that has a standard deviation of 78 s ($=0.00090$ d).
This is also an external approximate upper limit for our timing
precision, and agrees well with the uncertainties $\Delta T_0$ that we
estimated.

\begin{figure*}
\centering 
\includegraphics[width=9cm]{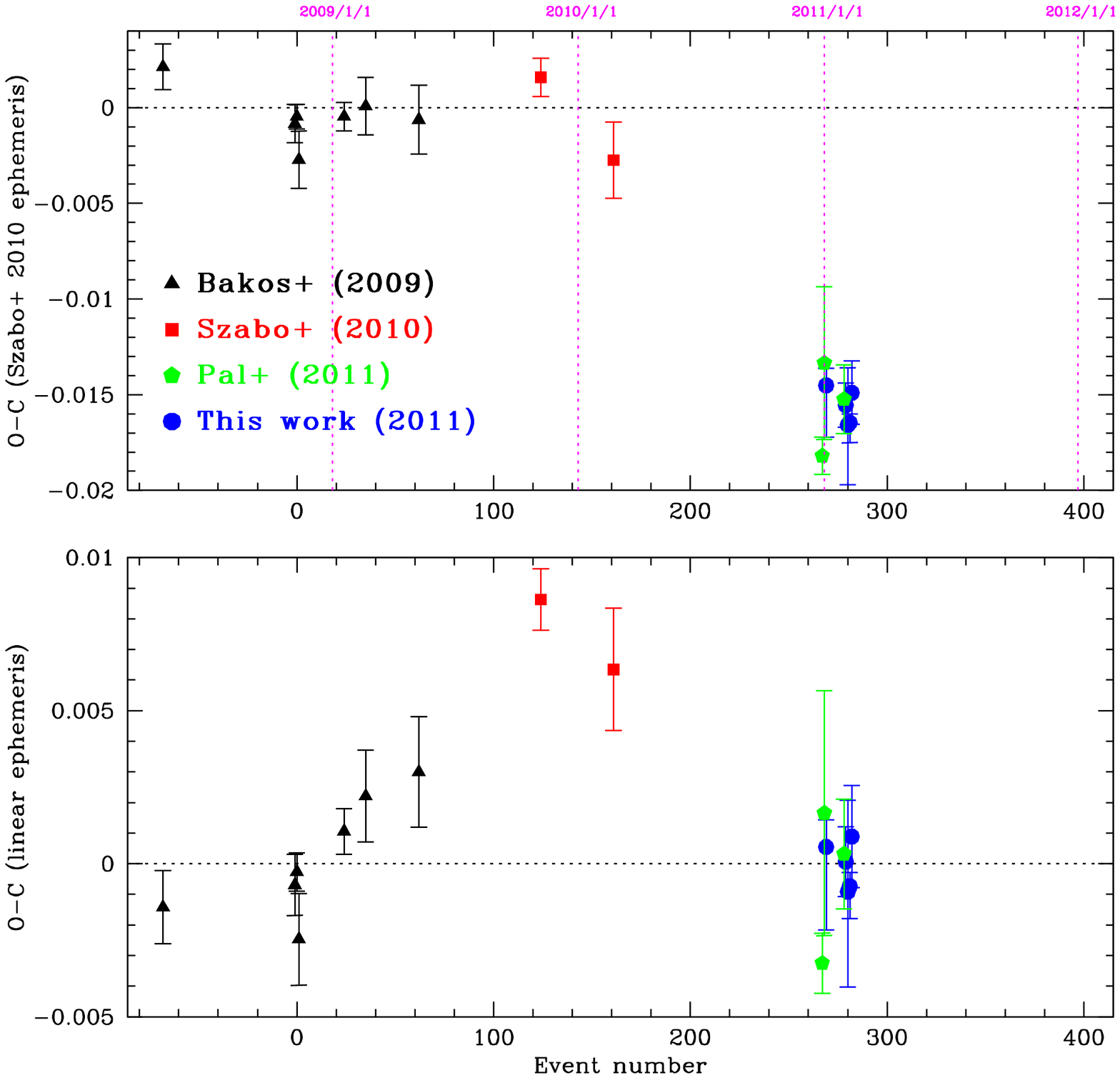}
\includegraphics[width=9cm]{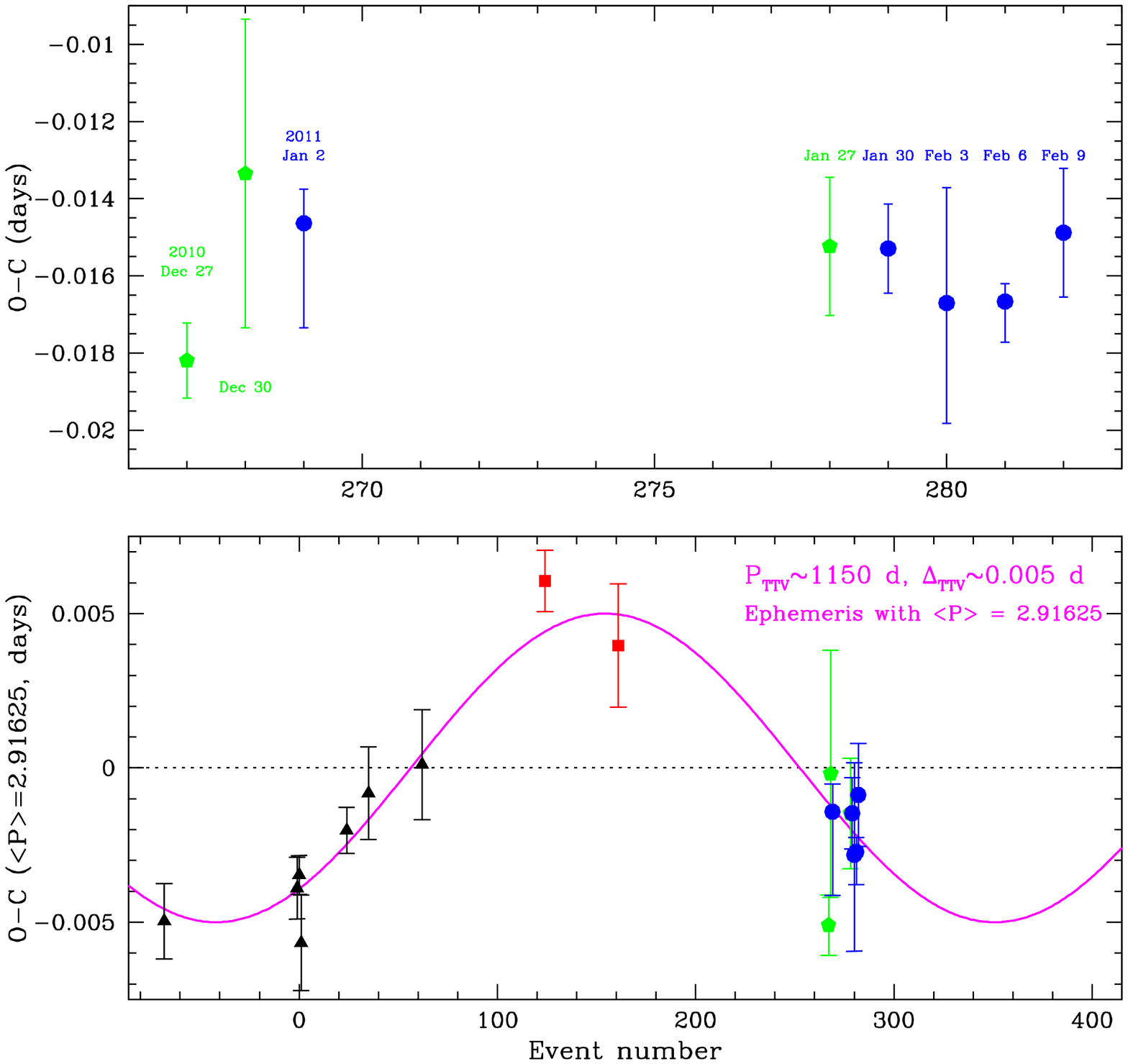}
\caption{\emph{Top left:} $O-C$ diagram following the \citet{pal2011} linear
ephemeris. The new points from TASTE (Table \ref{observ}) are plotted in blue
filled circles. 
\emph{Bottom left:} $O-C$ diagram following a linear ephemeris, fitted ignoring
the two data points (red squares) from \citet{szabo2010}.
\emph{Top right:} Same as top left, zoomed on the transits collected in Jan-Feb 2011.
\emph{Bottom right:} the $O-C$ diagram folded over an ephemeris with $\langle P\rangle=2.91625$ days,
perturbed by a sinusoidal
TTV with a period of $P_\mathrm{TTV}=1150$ days and an amplitude $\Delta_\mathrm{TTV}=0.005$ days.}
\label{ttvsinus}
\end{figure*}

In the top left panel of Fig. \ref{ttvsinus}, we plotted our
five new data points, along with the ones available from the literature
\citep{bakos2009,szabo2010,pal2011} in a $O-C$ diagram  using as a
reference  the ephemeris given by \citet{pal2011}. We confirm the
timings of \citet{pal2011}, with  additional, more precise
measurements.  Our transits collected in January/February 2011 lie,
respectively, 8.1$\sigma$, 13$\sigma$, 5.5$\sigma$, 22$\sigma$, and
8.9$\sigma$ from the linear ephemeris fitted  to the previous data. It
is clear that an updated linear ephemeris cannot be fitted to all the
available points with an acceptable $\chi ^2$. No  significant trend
in the $O-C$ diagram is visible for the 2011 transits
(Fig. \ref{ttvsinus}, top right panel). 

The two transits shown as red squares in the plots were observed under
non-optimal  weather conditions. During the first transit, ``sky was
photometric during the transit, but it was foggy in the evening and
from 40 min after the egress phase''. During the second, ``cirri were
present that significantly affected the $V$ band data, but the $R$
light curve was well reconstructed''  (from \citealt{szabo2010}).
Following a suggestion by  an anonymous referee, we checked whether a
linear ephemeris can be properly fitted by ignoring these two data
points (Fig. \ref{ttvsinus}, bottom left panel). 
%
All of the first four transits by \citet{bakos2009} lie at $O-C<0$ 
(two by more than 1$\sigma$), while the second three transits lie at $O-C>0$ 
(all by more than 1$\sigma$), suggesting a  systematic trend.  
In any case, it seems
unlikely that both the \citet{szabo2010} data points are outliers,  as
they deviate in the same direction by a consistent amount (12.4 min =
8.6$\sigma$, and 9.1 min = 3.2$\sigma$ respectively). They also come
from observations carried out 108 days apart, made with two different
telescopes by professional astronomers. As a cross-check, we also
tried to compare the timings presented by \citet{bakos2009},
\citet{szabo2010}, and \citet{pal2011} and by ourselves with data collected by
amateurs available from the Exoplanet Transit Database (ETD). None  of those
twenty-two light curves are reliable for our analysis, being plagued
to various extents by systematic errors: more than half of these
data points deviate by more than 1-$\sigma$ from a linear ephemeris.
Though this TTV needs to be confirmed in a future season, present
observational evidence points  towards an indication of an anomaly in
the periodicity of the transit.

We consider for a moment that the claimed TTV is real.
As this TTV appears as a sudden switch of the ephemeris,
\citet{pal2011} suggested that this deviation in the $O-C$ diagram
could be interpreted as a ``spike'' caused by a long-period eccentric
pertuber that is now near periastron. Examples of these systems
can be found in the synthetic $O-C$ diagrams plotted by
\citet{holman2005}. This could explain why the 2008-2010 timing points
are consistent with a linear ephemeris: such a perturber would have been
far from HAT-P-13b, and its perturbative effects well within the
measurement error. However, we note that the problem is still highly
unconstrained, owing to the large errors and the uneven sampling of
the previous measurements.  Follow-up observations are required to
constrain the mass and the orbit of the perturber without huge
degeneracies in the parameter space. In particular, we propose 1) to
search for any unpublished measurements performed in March--November
2010, when the rising part of the spike could have been sampled and 2) to
schedule new observations in October 2011--April 2012 , to check whether the
perturbation  is still active, or the new timing points  return to
the original mean ephemeris.

To demonstrate that the measurements are consistent with different
scenarios, we note that a periodic TTV cannot be excluded, in spite of
the conclusions of \citet{pal2011}. We folded the $O-C$
diagram over an ephemeris that had an average period $\langle P\rangle
=2.91625 $ days and had been perturbed by a sinusoidal TTV with an amplitude
$\Delta_\mathrm{TTV}=0.005$ days and a period $P_\mathrm{TTV}=1150$
days (Fig. \ref{ttvsinus}, last panel). This solution would be
perfectly consistent with the available data, with only one  outlier
(the 2010 Dec 27 transit by \citealt{pal2011}, $2.57\sigma$ from the
best-fit solution), and compatible with the presence of an outer,
coplanar, non-eccentric $5 M_{\oplus}$ perturber locked in a 3:2
mean-motion resonance with HAT-P-13b, following the analytical
approximations by \citet{agol2005}. This body  could  not have been
detected by the RV measurements carried out so far.  The observations
that will be taken in October 2011--April 2012 will allow us to discriminate
at least between this scenario and the ``eccentric perturber''
hypothesis, as a return to the original constant ephemeris would not
be compatible with a periodic perturbation.

\begin{acknowledgements}
We thank A. Cunial and M. Fiaschi for the night support
during the observations of HAT-P-13b. 
This work was partially supported by PRIN INAF 2008
"Environmental effects in the formation and evolution of extrasolar planetary
system". V.N. acknowledges support by STScI grant DDRF
D0001.82432.
\end{acknowledgements}

\bibliographystyle{aa}
\bibliography{16830}

\end{document}